\documentclass[aps,prb,twocolumn,superscriptaddress,noshowpacs,floatfix]{revtex4}
\usepackage{graphicx,epsfig}
\usepackage{amsmath}
\usepackage[breaklinks=true,colorlinks,citecolor=blue,linkcolor=blue,urlcolor=blue]{hyperref}
\def\be{\begin{equation}}
\def\ee{\end{equation}}
\def\nn{\nonumber}
\def\ber{\begin{eqnarray}}
\def\eer{\end{eqnarray}}
\begin{document}
\title{Finite time  interaction quench in a  Luttinger model}
\author{Rashi Sachdeva}
\affiliation{Department of Physics, Indian Institute of Technology, Kanpur 208016, India}
\author{Tanay Nag}
\affiliation{Department of Physics, Indian Institute of Technology, Kanpur 208016, India}
\author{Amit Agarwal}
\email{amitag@iitk.ac.in}
\affiliation{Department of Physics, Indian Institute of Technology, Kanpur 208016, India}
\author{Amit Dutta}
\affiliation{Department of Physics, Indian Institute of Technology, Kanpur 208016, India}

\begin{abstract}
We analyze the dynamics of a Luttinger model following a quench in the electron-electron interaction strength, where the change in the interaction strength occurs  over a 
finite time scale $\tau$. We study the Loschmidt echo  (the overlap between the initial and final state) as a function of time, both numerically and  within a perturbation scheme, treating the change in the interaction strength as a small parameter, for all $\tau$. We derive the corrections appearing in,  a.) the Loschmidt echo for a finite quench duration $\tau$, 
b.) the scaling of the echo following a sudden ($\tau \to 0$) quench, and c.) the scaling of the echo after an adiabatic ($\tau \to \infty$) quench. We study in detail, the limiting cases of the echo in the early time and infinite time limit,  and provide 
scaling arguments to understand these in a general context. We also show that our perturbative results  are in good agreement with the exact numerical ones.
\end{abstract}
%
%
\maketitle

\section{Introduction}

There is a recent upsurge in the studies of non-equilibrium dynamics of quantum many body systems
\cite{polkovnikov11,dutta10,dziarmaga10}
driven across a quantum critical point (QCP) or gapless critical regions \cite{sachdev99}. Possibility of experimental realization  in 
cold atomic systems \cite{coldatom_1,polkovnikov11} has paved the way for a plethora  of 
theoretical works to investigate the time dependent evolution 
and detection of quantum many body systems. In particular, 
quenching of interactions by means of Feshbach resonances 
or changing the lattice parameters as a function of 
time has motivated numerous theoretical \cite{polkovnikov11}
and experimental works \cite{coldatom_3}. 

In this article, we explore the behaviour of the Loschmidt echo (LE), 
which is defined as the overlap of wave functions, 
$|\Psi_{0}(t)\rangle$ and $|\Psi(t)\rangle$, evolving from the same initial state, 
but with different Hamiltonians $H_{0}$ and $H$, respectively.  It is given by
\be \mathcal{L}(t)=\left| \langle\Psi_{0}|e^{i H_0 t} e^{-i H t} |\Psi(t)\rangle \right|^{2},
\label{LE0}
\ee 
and is usually interpreted as a measure of the hyper-sensitivity of the time evolution of the system to the perturbations experienced due to the surrounding environment \cite{peres,scholar}.  It is also interpreted  as the time evolved fidelity
following a quantum quench \cite{pollmann10, heyl13} in the sense that $\mathcal{L}(t)$ measures the overlap between the initial ground state $|\Psi_0\rangle$ (of $H_0$) and the corresponding time evolved state $|\Psi(t) \rangle = e^{-i H t} |\Psi_0\rangle$, when $H_0$ is changed to $H$ . In the context of a quantum phase transition, the LE has been found useful in detecting
a QCP showing a sharp dip in its vicinity \cite{quan06,sharma12}; it also shows an early time decay with the decay constant characterized by
the critical exponents of the associated quantum phase transition \cite{fazio_pra}.  In recent years the LE, which is also related to the orthogonality catastrophe, has been 
probed experimentally \cite {Knapp_PRX}.
Although the temporal evolution  of the echo following a quantum quench across a
QCP has been studied in several  works \cite{silva08,pollmann10,damski11,Venuti,damski11_2,nag12,heyl13}, the same when the quantum system is quenched within a
gapless critical phase has gained more prominence in recent times  \cite{cazalila_prl06, Perfetto_epl, Dora_prl_2011,Dora_prl13, Dora3, Meden2, meden_prl12}.

In this work, we focus on a paradigmatic one dimensional interacting system with a gapless phase, namely  the Luttinger  model \cite{LL_ref} (LM) which is  characterized by bosonic
collective modes as elementary excitations. LM can be seen as a fixed point, in the renormalization group sense, for a large class of gapless quantum many-body systems in one dimension, {\it i.e.,} the equilibrium, low energy properties of many one-dimensional systems are universally described by the LM.
Interacting cold atoms 
in a one dimensional trap mimics such LMs \cite{coldatom_4}, as confirmed by existing 
experiments \cite{finitequench_1}. Other systems where
LM is relevant  are various spin models or interacting fermion
systems \cite{fazio_pra,dora_prb12,DzairmagaPRB2011}. 

The LM has recently been studied from the view point
of quantum quenches and thermalization \cite{cazalila_prl06, LL_quench2, aditi_prl11,aditi_prl12}. Here, we study the non-equilibrium properties of a LM, due to a change in the interaction parameter achieved over  finite span of time given by $\tau$, 
investigating the behaviour of the  LE.  
In particular we will focus on an interaction quench in a LM using a linear protocol from an initial state to a final state.

The paper is organized in the following manner. 
In Sec \ref{model}, we introduce the model Hamiltonian, {\it i.e.} LM,  with an interaction quench,  and
calculate the Loschmidt echo; within the central spin model, where a qubit is coupled to a quantum many body system, the LE measures the decoherence of the qubit \cite{quan06}. 
In Section \ref{adiasudsec}, we reproduce the limiting behaviour
of the LE for sudden ($\tau \to 0$) and adiabatic quench ($\tau \to \infty$), where we also argue that the result in 
the adiabatic case can also be interpreted by visualizing an adiabatic quench as a process that leads to the formation of an interacting Luttinger liquid   from a non interacting 1D system. 
In section \ref{quenchsec}, we study the LE for a finite time linear interaction quench  a.) within a perturbation scheme with
quench amplitude (or the change in the interaction strength) as the small parameter and b.) numerically.  In Section \ref{LElimits}, we analyze the  LE 
in various limits, particularly focussing on the early time (immediately after quench) and asymptotic ($t \to \infty$) behaviour
both for small and large $\tau$ and discuss alternate scaling arguments in a more general context.
The summary and discussion of our results is presented in section \ref{conclusions}. We note at the
outset that we shall denote the LE for fast quench (small $\tau$) and slow quench (large $\tau$ limit) using the notations $ \mathcal{L}_{SQ^{+}}$
and $ \mathcal{L}_{AQ^{-}}$, respectively.

\section{Luttinger model and the Loschmidt Echo}\label{model}

The low energy properties of interacting 1D bosons or that of spin chains 
 can be described in terms of bosonic sound like excitations, in the LM.
The initial  LM Hamiltonian, we consider is given by \cite{LL_ref}
\be
H_i=\sum_{q\neq 0} \left(v_0 |q|
b_q^\dagger b_q +\frac{g_i|q|}{2} \left[ b_q b_{-q} + b_q^\dagger b_{-q}^\dagger\right] \right) ~.
\label{Eq.1}
\ee
Here, $v_0 |q|$
is the `linearized' excitation spectrum of the non-interacting bosons, $g_i$ is the initial interaction strength, and $b_q^\dagger ~(b_q)$ is the creation (annihilation) operator describing the bosonic density excitations.
The  Hamiltonian in  Eq.\eqref{Eq.1} being quadratic in bosonic operators  can be easily diagonalized in terms of new bosonic quasiparticle operators ($\beta_q ,~ \beta_q^\dagger$), using the standard time-independent Bogoliubov transformation $b_q= u_q~\beta_q-v_q~\beta_{-q}^\dagger$. The commutation relations for the bosonic operators enforce the condition, $|u_q|^2 - |v_q|^2 = 1$, which enable us to
use the parameterization,  $u_q \equiv \cosh(\phi_q)$, and $v_q \equiv \sinh(\phi_q) $.
One can easily arrive at the condition that diagonalizes Hamiltonian (\ref{Eq.1}), given by,
\be
\tanh(2\phi_q)=\frac{g_i}{v_0}~ ~~{\rm or}~~~e^{-2\phi_q} = \left(\frac{v_0 - g_i}{v_0+g_i}\right)^{1/2} \equiv  K_i ~.
\ee
Here $K_i$ is the dimensionless LM interaction parameter which characterises the initial strength of interaction. Note that $K_i =1$ for a non interacting system, $K_i<1$ for repulsive ($g_i > 0$) electron electron interactions  and 
$K_i > 1$ for attractive ($g_i < 0$) electron electron interactions.
The Bogoliubov coefficients can be expressed as
\ber u_q &=& \frac{\sqrt{v_0 +g_i}+\sqrt{v_0-g_i}}{2 (v_0^2-g_i^2)^{1/4}} = \frac{K_i +1}{2\sqrt{K_i}}~,~~~{\rm and} \label{eq:u_q}\\
v_q &=& \frac{\sqrt{v_0 +g_i}-\sqrt{v_0 -g_i}}{2 (v_0^2 - g_i^2)^{1/4}} = \frac{K_i -1}{2\sqrt{K_i}} ~.
\eer
The diagonalised Hamiltonian in terms of new bosonic operators is given by 
\be \label{eq: GS}
H_i=\varepsilon_i +\sum_{q\neq 0} v_i |q| ~\beta_q^\dagger \beta_q,
\ee
where $v_i=\sqrt{v_0^2-g_i^2}$, is the renormalized velocity and $\varepsilon_i$ is the ground state energy of $H_i$ with respect to the non-interacting ground state.
The initial quasiparticle dispersion spectra is given by $\omega_i (q) =v_i |q|$; clearly, the ground state of $H_i$ is the vacuum of the $\beta$ bosons.

To study the dynamics of the LM, we quench the interaction strength from an initial value $g_i$ to a 
final value $g_f$ within a quench time $\tau$. 
The quench in the interaction parameter is described by  incorporating an additional time dependent term in the Hamiltonian [Eq.\eqref{Eq.1}]
\be
H'=\sum_{q\neq 0}\frac{\delta_g|q|Q(t)}{2}[b_qb_{-q}+b_q^\dagger b_{-q}^\dagger],
\ee
where $\delta_g =g_f-g_i$, and $Q(t)$ is the quench protocol satisfying $Q(t<0)=0$ and $Q(t>\tau)= 1$; the cases $\tau=0$ and $\tau \to \infty$ refer to the sudden and adiabatic quenching schemes, respectively.
The time dependent Hamiltonian ($H=H_i+H'$), can now be recast in terms of the $\beta$ bosons using the standard Bogoliubov transformation ($ b_q=u_q \beta_q-v_q\beta_{-q}^{\dag}$) to the form:
\ber
H(t)=\varepsilon_i+\sum_{q\neq 0}\left\{\left(v_i -\frac{g_i\delta_g Q(t)}{v_i}\right)|q|  \beta_q^\dagger  \beta_q+\right.\nonumber\\
\left.+\frac{\delta_g Q(t)v_0}{2 v_i}|q|[\beta_q \beta_{-q} + \beta_q^\dagger \beta_{-q}^\dagger ]
-\frac{\delta_g Q(t) g_i}{v_i}|q|\right\}~.
\label{Eq.3}
\eer 
Redefining the time-dependent parameters as
$v(t) = (v_i- \delta_g Q(t)  g_i/v_i)$, and 
$g(t) = (\delta_g Q(t)v_0/ v_i)$, and ignoring unimportant constants, one finds
\be
H(t)=\sum_{q\neq 0} \left(v(t) |q| ~\beta_q^\dagger \beta_q
+\frac{g (t) |q|}{2}\left[\beta_q \beta_{-q} + \beta_q^\dagger \beta_{-q}^\dagger\right] \right)\;.
\label{Eq.4}
\ee
The time evolution for the quadratic Eq.\eqref{Eq.4}
is obtained by the Heisenberg equation of motion, leading to
\ber
i \partial_t \beta_q &=& 
v(t) |q| ~\beta_q (t)+ g(t) |q| ~\beta_{-q}^\dagger(t) ~, \nonumber\\
i \partial_t \beta_{-q}^\dagger(t)  &=& - v(t) |q|  ~\beta_{-q}^\dagger(t)-  g(t) |q| ~\beta_q (t)~.
\label{Eq.EOM}
\eer
The coupled linear Eq.\eqref{Eq.EOM} have solutions of the following form
\ber
\beta_q(t)= f(q,t)~\beta_q(0) +h^*(q,t)~\beta^\dagger_{-q}(0)~,
\label{Eq.ansatz}
\eer
where the time-dependence has been completely shifted to the pre-factors $f(q,t)$ and $h(q,t)$ which satisfy the condition $|f(q,t)|^{2}-|h(q,t)|^{2} = 1$ for all times.  
Also,  the operators  $\beta_q(0)$ and $\beta_q^{\dagger} (0)$, appearing  on the right hand side of Eq.\eqref{Eq.ansatz} defined, at time
$t=0$, refer to non-interacting Bogoliubov bosons describing the initial Hamiltonian in  Eq.\eqref{eq: GS}.
Using Eqs.\eqref{Eq.EOM}-\eqref{Eq.ansatz}, we obtain coupled differential equations for the coefficients $f(q,t)$ and $h(q,t)$,  satisfying 
\be
i \partial_t \left[ \begin{array}{c}
f(q,t)  \\
h(q,t)
\end{array} \right] = |q| \times \left[ \begin{array}{cc}
v(t)  & g(t)    \\
-g(t)  & -v(t)
\end{array} \right]  
\left[ \begin{array}{c}
f(q,t)  \\
h(q,t)
\end{array} \right]~.
\label{Eq.6}
\ee
with the initial condition $[f(q,0),~h(q,0)] =[1,~0]$.

Using the generic definition  given in Eq.\eqref{LE0}, one can find the LE or the overlap of wave functions time evolved from the initial ground state  with the Hamiltonian $H_i$ and the time dependent  Hamiltonian $H(t)$ [Eq.~(\ref{Eq.4})], respectively. This  has been calculated for a spin-less Luttinger model in Ref. [\onlinecite{Dora_prl13}], and is expressed in terms of the time dependent coefficients $[f(q,t), h(q,t)]$ as, 
\ber
\mathcal{L}(t)=\exp {\left[-\sum\limits_{q>0}~\ln\left(|f(q,t)|^2\right)\right]}~.
\label{le}
 \eer
In subsequent sections, we solve Eq.\eqref{Eq.6}  
to obtain $[f(q,t), h(q,t)]$ and hence the LE in different situations.

\section{Loschmidt echo in the adiabatic and the sudden quench limit}\label{adiasudsec}
Before discussing the LE for a finite time quench, let us briefly reproduce the behaviour of the LE for the limiting cases of adiabatic and sudden quench reported in Ref.[\onlinecite{Dora_prl13}]. We reiterate that  we are varying the interaction strength from $g_i$ to $g_f$ in a finite interval of  time $\tau$; $\tau \to \infty$ ($\tau=0$) limit correspond to the adiabatic (sudden) case.
For convenience, let us also define the final renormalized velocity, $v_f = (v_0^2 - g_f^2)^{1/2}$, and the final dispersion relation, $\omega_f (q) = v_f |q|$.

\subsection{Adiabatic quench}
In the adiabatic limit, after the quench ($t > \tau \to \infty$, {\it i.e.}, $Q(t) =1$), Eq.\eqref{Eq.6} has stationary solutions of the form $[f(q,t), h(q,t)] = [f_q, h_q] e^{-i \omega_f(q) t} $. These solutions in Eq.\eqref{Eq.6} together with the final expression for the velocity and interaction strength, $v(t> \tau) = (v_0^2 - g_f g_i)/v_i$ and $g(t> \tau) = (g_f -g_i) v_0/v_i $,  leads to
\be 
|f(q,t>\tau)|^{2}=\frac{1}{2}+\frac{1}{4}\left(\frac{K_{f}}{K_{i}}+\frac{K_{i}}{K_{f}}\right)~.
\label{adiares}
\ee
Here we have used  the constraint $|f_q|^2-|h_q|^2 = 1$ and the parameter
$ K_{f}=\sqrt{[v_0-g_f]/[v_0+g_f]}$ characterizes
the strength of interaction in final state.  The LE can now be obtained by substituting Eq.\eqref{adiares} in Eq.\eqref{le}, and
is given by 
\be
\mathcal{L}_{AQ} =
\left(\frac{1}{2}+\frac{1}{4}\left(\frac{K_{f}}{K_{i}}+\frac{K_{i}}{K_{f}}\right)\right)^{-L/2\pi\alpha}~,
\label{ad1}
\ee
where  $L$ is the system size and the momentum sum in the exponential has been regularized using the ultraviolet cut-off $1/\alpha$. Here $\alpha$ is a non-universal and model-dependent short distance cutoff (inverse of the ultraviolet cutoff) which is used for the regularization of divergences in the Luttinger  model. 
Since $1/\alpha$ is a measure of the maximum wave vector included in the sum of Eq.{\eqref{le}}, and the separation between any two wave vectors is $ 2 \pi /L$, the exponent $L/(2 \pi \alpha)$, appearing in the Loschmidt Echo, just represents the number of wave vectors appearing in the sum of Eq.{\eqref{le}}.  The soft cutoff $\alpha$ arises, naturally when any finite size physical system is mapped to the Luttinger model. For example in the $XXZ$ model, 
$L/(2 \pi \alpha) = \pi^2 N \chi_{f}$, where $N$ is the number of lattice sites, and, $\chi_f$ is the known\cite{sirker_prl10}, fidelity susceptibility around the non-interacting $XX$ point of the $XXZ$ model \cite{Dora_prl13}. In the present study, the cut-off renormalizes the length of $L$ of the system
and the scaling relations derived here depend on the rescaled length $L/\alpha$.

Clearly the adiabatic LE simply measures the overlap of the ground state of 
initial and final Hamiltonians implying that in the adiabatic limit the system is always in its instantaneous ground state 
which can be viewed as the ground state of an  instantaneous LM  with a time dependent effective interaction parameter
\be
K(t) = \sqrt{\frac{v(t) - g(t)}{v(t) + g(t)}}~. \label{eq:Kt}
\ee
One can also use this argument  to arrive at the result given in Eq.~\ref{ad1}.  Using Eq.\eqref{eq:Kt} in Eq.\eqref{eq:u_q}, for $t > \tau$, {\it i.e.}, when $v(t) = (v_0^2 - g_f g_i)/v_i$ and $g(t) = (g_f -g_i) v_0/v_i $,  one can find out   the time independent Bogoliubov coefficient, 
\be
|u_q|^2 = \frac{1}{2}+\frac{1}{4}\left(\frac{K_{f}}{K_{i}}+\frac{K_{i}}{K_{f}}\right)~. \label{ad2} 
\ee
 After the quench, $|u_q|^2 = |f(q,t>\tau)|^2$ as the interaction parameter is no longer changing and there is no mixing of the $\pm q$ modes [see Eq.\eqref{Eq.ansatz}], and this leads to Eq.\eqref{ad1}. 
 
We note in passing that the argument presented above is also consistent with the idea behind the `formation' of an interacting Luttinger liquid state starting from an initial non-interacting state, $K_i = 1$, by adiabatically switching  on the interactions. Technically this is evident by substituting $K_i \to 1$ and $K_f \to K_i$ in Eq.\eqref{ad2} which immediately leads to Eq.\eqref{eq:u_q}.  
 We note that the adiabatic LE turns out to be the modulus squared of the ground state fidelity, and Eq.\eqref{ad1} is consistent with the ground state fidelity calculated earlier \cite{fidelity_yang}. 

\subsection{Sudden quench}
In the sudden quench limit ($\tau \to 0$), for $t> 0$, the velocity and the interaction strength assume their
time independent final value, and Eq.\eqref{Eq.6} can be decoupled to obtain $\partial_t^2 f (q,t) + v_f^2 |q|^2 f (q,t) = 0$. 
Solving this with the initial condition, $f(q,0) =1$,  we get
\be
f(q,t)=\cos(v_{f}|q|t)-\frac{i}{2}\left(\frac{K_{f}}{K_{i}}+\frac{K_{i}}{K_{f}}\right)\sin(v_{f}|q|t) ~.
\label{eq:SU1}
\ee
(An identical result has also been derived in Ref.[\onlinecite{cazalila_prl06}] using forward Bogoliubov transformation which is followed by the backward one). Using Eq.\eqref{eq:SU1}, we find  the decay in the LE  in the very short
time limit ($t \ll \alpha/v_{f}$)) given by
\be
\mathcal{L}_{SQ}(t)
\approx \exp\left[-\frac{L}{2 \pi \alpha}\left(\frac{K_f}{K_i} - \frac{K_i}{K_f} \right)^2 \frac{v_f^2 t^2}{12 \alpha^2}\right]~.\label{sud1} 
\ee
We note that this is the usual gaussian decay of a LE expected (from a perturbation theory point of view) for a sudden quench \cite{peres, fazio_pra}. 

In the large time limit ($t\gg\alpha/v_f$), on the other hand,  one can neglect a small oscillating term with decaying amplitude (as $t$ increases), to obtain  a simplified time independent form of the echo, 
\be
\mathcal{L}_{SQ}(t\gg\alpha/v_f) =
\left(\frac{1}{2}+\frac{1}{4}\left(\frac{K_{f}}{K_{i}}+\frac{K_{i}}{K_{f}}\right)\right)^{-L/\pi\alpha}~.
\label{suddenlong}
\ee

We emphasize  that in the long time $t\rightarrow\infty$ limit, $\mathcal{L}_{SQ}=\mathcal{L}_{AQ}^2$; this  correspondence has been reported  in the Ref. [\onlinecite{Dora_prl13}] (also see Ref. [\onlinecite{Heyl13}]). 
In the adiabatic limit,    $\mathcal{L}_{AQ}=|\langle G_f|G_0\rangle|^2$ where $G_0$ and $G_f$ are the ground states of the initial and final Hamiltonian, respectively.
On the other hand,  in the sudden limit, the  LE takes the form $
\mathcal{L}=|\langle\Phi(t)|\Phi_0(t)\rangle|^2 = |\langle G_0|e^{i H t} e^{-i H_0 t} | G_0\rangle|^2$.
When expanded in eigenstates
of final Hamiltonian, only the ground state contributes to LE in the asymptotic limit, while the oscillating terms due to
the excited states interfere destructively and vanish as $t\rightarrow\infty$; one therefore finds  $\mathcal{L}_{SQ}|_{t \to \infty}=|\langle G_f|G_0\rangle|^4$, thereby yielding the correspondence $\mathcal{L}_{SQ}=\mathcal{L}_{AQ}^2$. 

\section{Loschmidt Echo for the finite time linear quench}\label{quenchsec}

In this section we focus on the interaction quench in a LM from an initial value $g_i$ to a final
value $g_f$ introduced over a finite interval time $\tau$ using the quenching protocol,
$$Q(t\le0)=0, ~~~Q(0<t<\tau)=t/\tau ~~{\rm and}~~ Q(t\ge\tau)= 1.$$
We use Eq.\eqref{Eq.6} which can be decoupled in different regimes.  For all  quenching
schemes, we get
\be
\ddot{f} + v_i^2q^2 f =  0 ~~~{\rm and} ~~~
\ddot{h} + v_i^2q^2 h =0~,
\ee
for $t<0$ while  for $t \ge \tau$ \be
\ddot{f} + v_f^2q^2 f =  0 ~~~{\rm and} ~~~
\ddot{h} + v_f^2q^2 h =0~.
\label{Eq.21}
\ee
where we have dropped the arguments $q,t$ for $f(q,t)$ and $h(q,t)$.
However for the quenching scheme we are interested in, we have the following set of coupled equations during the interval  $0< t < \tau$~ {\it i.e.}, when the interaction is being quenched
 \ber
\label{Eq.9}
 \ddot{f} - \frac{\dot{f}}{t} + \left[v_0^2q^2 -  \left(g_i + \delta_g t/\tau\right)^2q^2 - i \frac{v_i|q|}{t} \right]f &=&  0 ~, \nn \\
 \ddot{h} - \frac{\dot{h}}{t} + \left[v_0^2q^2 - \left(g_i + \delta_g t/\tau\right)^2q^2
 + i \frac{v_i|q|}{t} \right]h &=&0 ~. \nn \\
\label{Eq.linear}
\eer
We solve these linear second order inhomogeneous equation using numerical techniques.
However, we also consider the case when the change in the  interaction parameter 
$\delta_g$ is small and employ a perturbative expansion in $\delta_g \to 0$,  
to gain  insight of the time-evolution of the LE following a linear quench. We emphasize that the $\delta_g \to 0$ limit happens to be analogous to the adiabatic quench regime of $\tau \to \infty$, as only the combination of $\delta_g/\tau$ appears in Eq.\eqref{Eq.9}. 

We now find the perturbative solutions of Eq.\eqref{Eq.6} [or equivalently Eq.\eqref{Eq.9}], with the boundary conditions $[f(q,t=0),h(q,t=0)] =[1,0]$, in terms of the small interaction parameter, {\it i.e.}, $ \delta_g << v_i$. The solutions of Eq.\eqref{Eq.21} are just harmonically varying functions of constant frequency ($\omega_f$), whose constant coefficients need to be determined through boundary condition at $t = \tau$; this necessitates solving Eq.\eqref{Eq.9} to obtain the value of the functions $[f(q,\tau),h(q,\tau)]$.

{\bf Solution for $t < \tau$} ({\it i.e.}, within the interval  during which the interaction term is being changed): using Eq.\eqref{Eq.6},  we get   \cite{Dora_prl_2011} to lowest order in $\delta_{g}$, {\it i.e.}, ${\cal O}(\delta_g)$, \ber
f(q,0<t<\tau)&\approx& \text{exp}(-iv_{i}|q|t)\label{uqpert} \nonumber\\
h(q,0<t<\tau)&\approx&\frac{i\delta_{g}v_{0}|q|}{v_{i}\tau}~
e^{i[v_i t-\frac{\delta_{g}g_{i}}{2v_{i}\tau}t^{2}]|q|} \nonumber \\ 
& \times & \int_{0}^{t}t'e^{-2iv_{i}|q|t'}e^{i\frac{\delta_{g}g_{i}|q|}{2v_{i}\tau}t'^{2}}dt'~.
\label{hqtfull}
\eer
The integral on the RHS of above equation can be evaluated exactly to obtain, 
\begin{widetext}
\ber h(q,t)& =&\frac{v_{0}}{g_{i}}~    
e^{-i\left(2 v_{i}^{2}t+\frac{g_{i}\delta_{g}t^{2}}{\tau}
+\frac{4v_i^{4}\tau}{g_{i}\delta_{g}}\right)\frac{|q|}{2v_{i}}}
 \Bigg[e^{\frac{2i v_{i}^{3}\tau |q|}{g_{i}\delta_{g}}}
\left(-e^{2iv_{i}|q|t}+e^{\frac{ig_{i}\delta_{g}|q|t^{2}}{2v_{i}\tau}}
\right) +(1+i) ~e^{2iv_{i}|q|t}  \nonumber\\
&\times & v_{i} \sqrt{\frac{\pi\tau v_{i}|q|}{g_i\delta_g}} 
 \Bigg\{ \mathbf{Erfi}\left(\frac{(1+i)v_{i}
 \sqrt{v_{i} |q| \tau}}{\sqrt{g_{i} \delta_g}}\right)+\mathbf{Erfi} 
 \left(\frac{ (1+i)(g_{i}\delta_{g}t-2v_{i}^{2}\tau)
\sqrt{|q|}}{2 \sqrt{g_{i}\delta_{g}v_{i}\tau}}\right)\Bigg\}\Bigg],~\label{hqttletau} 
\eer
\end{widetext}
where $\mathbf{Erfi}(z) = -i ~\mathbf{Erf}(i z)$,  is the imaginary error function.
We expand the expression in Eq.(\ref{hqttletau}) in powers of $\delta_{g}$,
retaining only the lowest order term.
This simplifies Eq.\eqref{hqttletau} upto ${\cal O}(\delta_g)$ to the form, 
\be
h(q,t\le\tau)\approx \frac{\delta_{g}v_{0}}{2v_{i}^{2}} \left( \frac{\sin(v_{i}|q| t)}{v_i |q| \tau}- \frac{t}{\tau}e^{-iv_{i} |q|t} \right)
~. \label{hqtle2}\ee

{\bf Solution for $t\ge\tau$} : In order to calculate the coefficients 
in $t\ge\tau$ regime, {\it i.e.}, after the interaction reaches its final value, we use Eq. \eqref{hqtle2} at $t=\tau$ as the boundary condition for Eq. \eqref{Eq.21} and obtain the time evolution of $h(q,t\ge\tau)$ to be
\be
h(q,t\ge\tau) \approx\frac{\delta_{g}v_{0}}{2v_{i}^{2}}\left[\frac{\sin(v_{i}|q|
\tau)}{v_{i}|q|\tau} e^{iv_{i}|q|(t-\tau)} - e^{-iv_{i}|q|t}\right] ~.
\label{fullhqtpert}
\ee
The first term in the square brackets dominates the time evolution for small $\tau$ ({\it i.e.}, $\tau << 1/\omega_i$), and the second term yields the time evolution for large $\tau$ ({\it i.e.}, $\tau >> 1/\omega_i$). 
One can readily obtain the exact results for the sudden ($\tau = 0$) as well as  adiabatic ($\tau = \infty$) quench limits 
\ber
h_{SQ}(q,t\ge\tau)& \approx&\frac{i\delta_{g}v_{0}}{v_{i}^2} \sin(v_{i}|q|t)~, ~~~{\rm and}\nonumber\\
 h_{AQ}(q,t\ge \tau)&\approx&-\frac{\delta_{g}v_{0}}{2v_{i}^{2}} e^{-i v_i|q| t }~,
\eer
to lowest order in $\delta_g$.
Eq.\eqref{fullhqtpert} provides the perturbative solution of $h(q,t)$ (which in fact yields the 
 number of excited states with respect to the 
vacuum of the $\beta_q$ bosons generated during the quenching process)
 in powers of $\delta_g$,  for finite $\tau$ at all times and appropriately reduces to the  sudden and adiabatic quench limits. 

We now proceed  to study the  LE by means of the perturbative solution in powers of small parameter $\delta_g$. 
Using Eq.\eqref{fullhqtpert}, and the constraint $|f(q,t)|^2 -|h(q,t)|^2 =1$, we find $|f(q,t)|^2$, which is then substituted in Eq.\eqref{le} to  obtain the 
finite $\tau$  behaviour of LE for a finite time, {\it i.e.,}  $t\ge\tau$ in the perturbative limit:

\begin{widetext}
\ber
\mathcal{L}(t)
&=&\text{exp}\left[-\frac{L}{2\pi}\int_{0}^{1/\alpha} 
\text{ln}\left( 1+\frac{\delta_{g}^{2}v_{0}^{2}}
{8v_{i}^{4}}\left[\frac{1}{v_i^2q^2\tau^2}+2-\frac{\cos(2v_{i}|q|
\tau)}{v_i^2q^2\tau^2}+\frac{2}{v_{i}|q|\tau}\{\sin(2v_{i}|q|
(t-\tau))-\sin(2v_{i}|q|t)\}\right]\right) dq \right]~, \nonumber\\
&\approx& \text{exp}\left[ -\frac{L}{2\pi \alpha}\frac{\delta_g^{2}
v_{0}^{2}}{4 v_{i}^{4}}\left(1-\frac{\alpha^2}{v_i^2\tau^2}\sin^2\left(\frac{v_{i}\tau}{\alpha}\right)+
\frac{ \alpha}{v_{i}
\tau}\left\{-\text{Si}\left(\frac{2v_{i}t}{\alpha}\right)+\text{Si}\left(\frac{2v_{i}(t-\tau)}{\alpha}\right)+\text{Si}\left(\frac{2v_{i}\tau}{\alpha}\right)\right\}\right)\right]~.\nonumber\\
\label{LEexp1}\eer
\end{widetext}
In the above equation, the function $\text{Si}(x)$ is the SinIntegral of $x$ defined as $\text{Si}(x)=\int_{0}^{x}(\sin{t}/t)dt$, 
with $1/\alpha$ as the upper momentum (ultra-violet) cut-off. 
The above expression provides a generic form of the LE as a function of  $\tau$ and $t$ which  will be used extensively in the subsequent calculations.
In the adiabatic quench case ($\tau \to \infty$), only the first term in the exponential of Eq.\eqref{LEexp1} contributes, while in the sudden quench limit ($\tau \to 0$), 
the first and the second terms in the exponential of Eq.\eqref{LEexp1} contribute, resulting in the following expressions, 
\ber
 \mathcal{L}_{AQ}(t \ge \tau)&=& \text{exp} \left[-\frac{L}{2\pi\alpha}\frac{\delta_{g}^{2} v_0^{2}}{4v_{i}^{4}}\right] ~, ~~~{\rm and} \label{LEpertAQ1} \\
 \mathcal{L}_{SQ}(t \ge \tau)&=& \text{exp} \left[-\frac{L}{2\pi\alpha}\frac{\delta_{g}^{2} v_0^{2}}{2v_{i}^{4}}\right] ~ . \label{LEpertSQ1} 
\eer
which satisfy the correspondence $\mathcal{L}_{SQ}=\mathcal{L}_{AQ}^2$ established for an arbitrary
value of $\delta_g$ \cite{Dora_prl13}.

In Fig.{\ref{LEplot1}}, we plot the temporal evolution of  the LE  for a fast quench, with $\tau = 0.2$, obtained
both numerically and analytically for small $\delta_g$ . As anticipated earlier, we find very good agreement between the perturbative and the exact results.

\begin{figure}[t]
\includegraphics[width=0.5\textwidth]{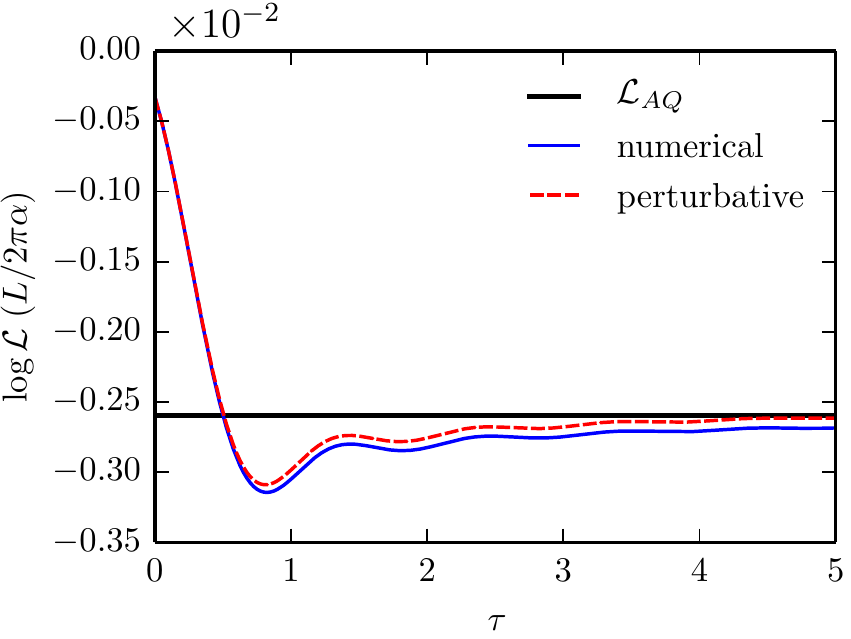}
\caption{(Color online) Plot for $\log \mathcal{L}(t)$, in units of $L/(2 \pi \alpha)$,  in the early time limit as a function of $\tau$. 
Here we have chosen $t = \tau + 0.1$, and $\delta_{g}/v_i=0.1$ for an initial choice of $g_i/v_i = 0.1$.
The solid (blue) line denotes the exact numerical result and the dashed(red) lines shows the perturbative solution. 
The solid horizontal line is the analytical expression for the LE in the sudden quench limit (for $t \to \infty$). 
}
\label{LEplot3}
\end{figure}

The case of a slow quench, for $\tau = 5$, is shown in Fig.{\ref{LEplot2}. Although, the quantitative agreement between the perturbative and the exact solutions is not as
perfect as the small $\tau$ case, the perturbative solutions still capture all the qualitative features. However, it is worth noting that for both fast and slow quenches, the LE shows a damped oscillatory behaviour which saturates to some finite value in the infinite time limit. The dimensionless time period of these oscillations can be estimated using Eq.\eqref{LEexp1} and is given by $T =  \pi \alpha/(v_i \tau)$, which corresponds to a value of $5$ and $0.2$ for Fig.\ref{LEplot1} and Fig.\ref{LEplot2} respectively. The damped oscillatory nature of the LE, and its saturation to
an asymptotic constant value with time, are characteristics of the LE that has also been observed  following a linear quench across the  QCP of a transverse Ising chain \cite{pollmann10}.

\begin{figure}[t]
\includegraphics[width=0.5\textwidth]{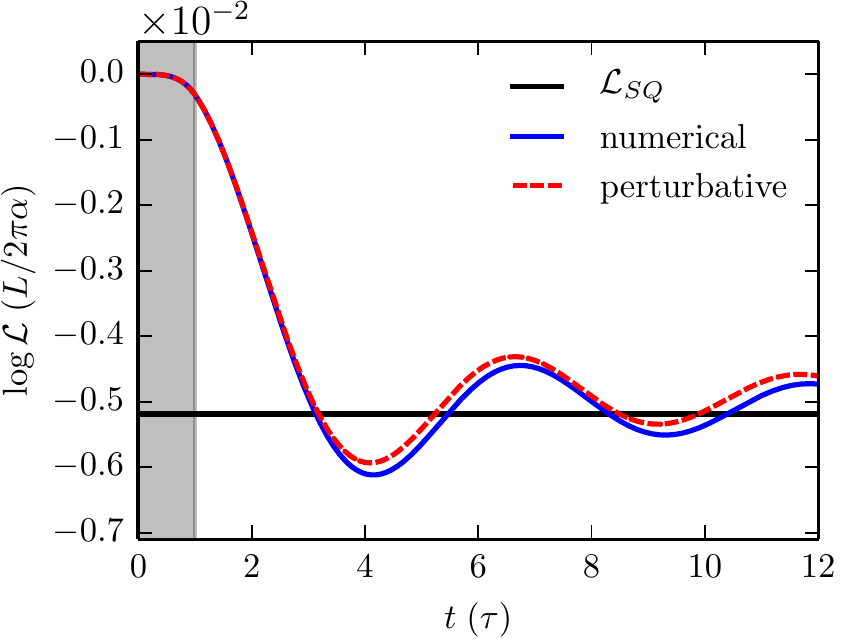}
\caption{(Color online) Plot for $\log \mathcal{L}(t)$, in units of $L/(2 \pi \alpha)$, as a function 
of time for $v_i \tau/\alpha = 0.2\pi$ and  $\delta_{g}/v_i=0.1$ for an initial choice of $g_i/v_i = 0.1$ ({\it i.e.}, $K_i = 0.91$). The solid (blue) line denotes the exact numerical result [based on Eq.\eqref{Eq.6}] and the dotted(red) line shows the perturbative solution [based on Eq.\eqref{LEexp1}]. The solid horizontal line is the analytical expression for the LE in the sudden quench limit (for $\tau \to 0$ and $t \to \infty$). The shaded region marks the `quench' interval during which the interaction parameter is changed.}
\label{LEplot1}
\end{figure}

\begin{figure}[t]
\includegraphics[width=0.5\textwidth]{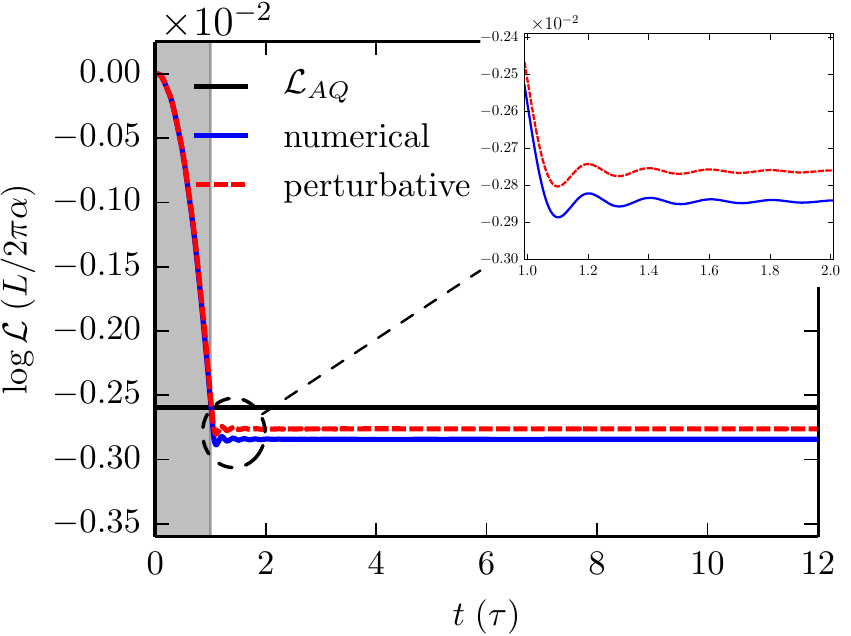}
\caption{ (Color online) Plot for $\log \mathcal{L}(t)$, in units of $L/(2 \pi \alpha)$, as a function 
of time for $v_i \tau/\alpha = 5\pi$ and  $\delta_{g}/v_i=0.1$ for an initial choice of $g_i/v_i = 0.1$ ({\it i.e.}, $K_i = 0.91$). The solid (blue) line denotes the exact numerical result [based on Eq.\eqref{Eq.6}] and the dotted(red) lines shows the perturbative solution [based on Eq.\eqref{LEexp1}]. The solid horizontal line is the analytical expression for the LE in the adiabatic quench limit (for $t \to \infty$). The shaded region marks the `quench' interval during which the interaction parameter is changed.}
\label{LEplot2}
\end{figure}
\section{Loschmidt Echo in different limits}\label{LElimits}
In this section we focus our efforts on studying the behaviour of the LE in the early time limit immediately after the quench [$(t/\tau-1) \to 0$] and also the asymptotic large time limit [$t/\tau \to \infty$]. In particular, we will be interested in the corrections to the LE in the sudden and the adiabatic limits.

\subsection{Loschmidt Echo in the early time limit}\label{LEtau}
To find the early time limit after the quench, {\it i.e.}, when $ 0 < t-\tau < \alpha  /v_{i}$, Eq.\eqref{LEexp1} can be expanded in powers of $t - \tau$ (see Fig. \ref{LEplot3}). 
For the case of  $t = \tau$,  the last term in Eq.\eqref{LEexp1} vanishes and 
LE  simplifies to the form
\be \mathcal{L}(\tau)= \exp
\left[-\frac{L}{2\pi\alpha}\frac{\delta_{g}^{2}v_0^{2}}{4v_{i}^{4}}
\left(1-\frac{\text{sin}^2\left(v_{i}\tau/\alpha\right)}
{v_{i}^2\tau^2/\alpha^2}\right)\right].~
\ee

The behaviour of the early time LE for a fast quench, {\it i.e.}, in the small $\tau$ limit, can be investigated by taking
the limit $\tau \to 0$ in Eq. \eqref{LEexp1}, and to the lowest order in $\tau$ for $ 0 < t-\tau < \alpha  /v_{i}$, it is given by
\be
\mathcal{L}_{SQ^{+}}({t}) \approx\text{exp}\left[-\frac{L}{2\pi\alpha} \frac{\delta_g^2 v_0^2}{3 \alpha^2 v_i^2} 
\left\{t^2-
\tau(t -\tau/4) \right \}\right].
\label{sqttLE} 
\ee
Here the first term corresponds to the expected gaussian decay of the LE, with a decay constant independent of $\tau$, which
is a generic feature of the LE following a quench \cite{peres, fazio_pra}.  On the other hand, the second term depends on $\tau$, 
and indicates a correction to the proper Gaussian 
decay;  thereby it carries a signature of the fact that the interaction has been quenched over
a finite interval of time. As expected, this correction term vanishes when $\tau=0$.}

For a slow quench, {\it i.e.}, large $\tau$, to lowest order in $\tau$ we find the following form for the LE for all times,
\ber
\mathcal{L}_{AQ^{-}}(t)&=& \text{exp}
\Bigg[-\frac{L}{2\pi\alpha}\frac{\delta_{g}^{2}
v_0^{2}}{4v_{i}^{4}}\Bigg\{1-\frac{\text{sin}^2(v_{i} \tau/\alpha)}{v_{i}^2\tau^2/\alpha^2} \nonumber \\
&+& \frac{2(t-\tau)}{\tau}-\frac{2v_{i}^{2}}{\alpha^{2}}\frac{(t-\tau)^{2}}{\tau}\frac{\text{cos}(2v_{i}\tau/\alpha)}{2v_{i}\tau/\alpha}\Bigg\}\Bigg]~. 
\nonumber \\
\label{eq:slow1}
\eer
Here the first term in the exponential,  is the echo for the adiabatic case and as expected it is independent of time and the quench rate.
The expression in Eq.\eqref{eq:slow1},  can be simplified to obtain the early time behaviour for a slow quench.  Retaining  the most dominant correction term in $\tau$ for $ 0 < t-\tau < \alpha  /v_{i}$, we obtain:
\be 
\mathcal{L}_{AQ^{-}}( {t })\approx\text{exp}
\left[-\frac{L}{2\pi\alpha} \frac{\delta_g^2 v_0^2}{ v_i^4} \left\{ 1+
2 \frac{ (t-\tau)}{\tau} \right \}\right]~.
\label{aqtfLE} 
\ee
We emphasise that, for a finite quench rate, we find that the correction to the early time behaviour of the LE, after a slow quench, shows a linear exponential decay in time. 

The corrections to the early time behaviour of the LE due to a finite quench rate, primarily arise due to the fact that the state from which the early time behaviour is observed, {\it i.e.} $\psi(t = \tau)$, is neither a ground state of the initial Hamiltonian, nor of the final Hamiltonian. In general $\psi(t = \tau)$  incorporates `defects' (excited states contributions) for a finite quench rate ($\tau$), over the initial ground state at $t=0$. This actually leads to the difference in the early time behaviour both in the large $\tau$ and small $\tau$ limits, manifested in Eqs. \eqref{sqtfLE} -\eqref{aqtfLE}.  

\subsection{Loschmidt Echo in the large time limit}\label{LEinf}
In this subsection, we use the exact expression of LE [Eq.\eqref{LEexp1}], in the perturbative limit of $\delta_g$, to study the behaviour of LE in the infinite time limit. As  $t \to \infty$,  Eq.\eqref{LEexp1} reduces to, 
\ber
\mathcal{L}(t \to \infty) &\approx& \text{exp}\Bigg[ -\frac{L}{2\pi \alpha}\frac{\delta_g^{2}
v_{0}^{2}}{4 v_{i}^{4}}\Bigg(1-\frac{\text{sin}^2\left(v_{i}\tau/\alpha\right)}
{v_{i}^2\tau^2/\alpha^2}\nonumber\\
&& + \frac{ \alpha}{v_{i} \tau}\text{Si}\left(\frac{2v_{i}\tau}{\alpha}\right)\Bigg)\Bigg]~.
\label{LEinf1}
\eer

\begin{figure}[t]
\includegraphics[width=0.5\textwidth]{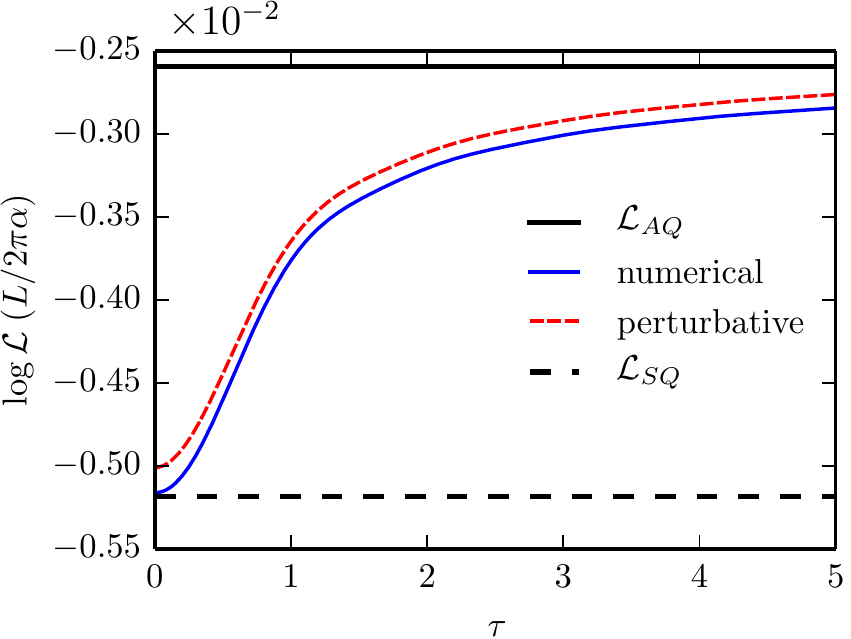}
\caption{Plot for $\log \mathcal{L}(t)$, in units of $L/(2 \pi \alpha)$,  in the large time (asymptotic) limit as a function of $\tau$. 
Here we have chosen $t = \tau + 50$, and $\delta_{g}/v_i=0.1$ for an initial choice of $g_i/v_i = 0.1$.
The solid (blue) line denotes the exact numerical result and the dashed(red) lines shows the perturbative solution. 
The  solid (dashed) horizontal line is the analytical expression for the LE in the adiabatic (sudden) quench limit for $t \to \infty$. 
}
\label{LEplot4}
\end{figure}

For small $\tau$, incorporating  correction to the  lowest order in $\tau$ appearing  in Eq.\eqref{LEinf1}, we find
\ber \mathcal{L}_{SQ^{+}}(t \to \infty) \approx\text{exp}\left[-\frac{L}{2\pi\alpha} \frac{\delta_g^2 v_0^2}{2 v_i^4} \left\{ 1-
\frac{v_i^2 \tau^2}{18 \alpha^2} \right \}\right]; 
\label{sqtfLE} \eer
where  the sudden quench result [Eq.\eqref{LEpertSQ1}] is recovered 
for $\tau = 0$. This is to be noted that the lowest order correction term (over the $\tau=0$ case) scales as $\tau^2$ which
can be be understood physically using a simple quantum mechanical argument.
The system changes over a time scale of  $\tau$, and the time evolved wave function can be obtained by integrating the  Schr\"odinger equation within the interval $[0,\tau]$:
 \be
 |\psi (\tau) \rangle - |\psi(0) \rangle = -\frac {i}{\hbar} \int_0^{\tau} H(t) ~dt~ . 
 \ee
 Since the integrand is finite, the integral is of the order of $\tau$ [see Ref.~\onlinecite{shankar}].
 We therefore get a correction which varies as  $\tau^2$, in the probability of excitation to the $n$-th excited state, which is given by  $P_n = |\langle\psi_0 | \psi_n \rangle|^2$.  
 Here $|{\psi_0}\rangle$ is the initial ground state and $|{\psi_n}\rangle$ is the $n$-th excited state associated with the final time evolved Hamiltonian. This correction scaling as $\tau^2$ appears in the  LE when LM is being quenched  with a small but finite $\tau$. 

In the limit of large $\tau$, the  LE in the asymptotic limit  is obtained by retaining only lowest order term
(of the order $1/\tau$) in Eq.\eqref{LEinf1}; this is given by

\ber
\mathcal{L}_{AQ^{-}}(t \to \infty)\approx\text{exp}\left[-\frac{L}{2\pi\alpha}\frac{\delta_g^2 v_0^2}{4 v_i^4} \left\{ 1+
\frac{\pi \alpha}{2\tau v_i} \right \}\right].
\label{aqtfLE}
\eer
The second term of Eq.(\ref{aqtfLE}) represents the first order correction in $\tau$ over the adiabatic quench ($\tau=\infty$ limit). We emphasise 
that the correction to the LE  over the adiabatic  limit for $t \to \infty$ scales as $1/\tau$ .

This can be understood by the following argument \cite{Grandi}. Lets assume a parameter $\lambda$ of a $d$-dimensional Hamiltonian  is driven as  $\lambda(t)=t/\tau$ within the gapless phase with quasiparticle energy dispersion as $\epsilon_q \sim A(\lambda)q^z$, where $z$ is the dynamical critical exponent. 
When the dynamics   is adiabatic,  the excitation to higher energy state becomes suppressed when the rate of change of $\lambda$ is small compared to the internal time scale, {\it i. e.},  $\dot \epsilon_q(\lambda)/\epsilon_q(\lambda) \ll \epsilon_q(\lambda)$. For the present model,  the inherent time scale is given by $L/v_i$ while the external time scale is $\tau$. Adiabaticity breaks  down when $\dot \epsilon_q(\lambda) \sim \epsilon_q^2(\lambda)$.  This leads to a characteristic momentum scale which is related to the quench rate $\tau$ through
the following relation:
\be
\tilde{q}^z\sim \frac{1}{\tau A(\lambda)}\frac{\delta {\ln A(\lambda)}}{\delta \lambda}.
\label{scale}
\ee 
The total number of quasiparticle excitation to higher state is proportional to the phase space volume ${ \tilde q}^d \sim \tau^{-d / z}$. This scaling law holds true for $d/z \le 2$. In the present model, therefore the measure of the defect density  given by  $|h(q,t)|^2$ integrated over all momenta mode  scales as $1/\tau$; this scaling is reflected in the asymptotic behaviour of the LE.

 One can observe that $ \mathcal{L}^{SQ}_{t \to \infty}=(\mathcal{L}^{AQ}_{t \to \infty})^2$ with the proper adiabatic and sudden limits of $\tau$ i.e., $\tau=\infty$ for perfect adiabatic quenching and $\tau=0$ for perfect sudden quenching. But when LM is being quenched through a finite driving rate $1/\tau$, the infinite time expressions of LE, for a fast and a slow quenching scheme, is modified in a relevant way.
 The ground state conjecture does not hold true for a finite $\tau$. The LE in the slow quenching case could not be described as an modulo square overlap between initial and final ground state of LM. The finite $\tau$ brings additional contributions coming from higher excited state to the infinite time LE. In the infinite time limit LE under a fast quenching scheme is also modified from the $\tau=0$ limit of LE.

\section{conclusion}
\label{conclusions}

In this manuscript, we have studied  the dynamics of a Luttinger model following an  interaction quench when the quench is applied
over a finite  duration $\tau$.  The Loschmidt echo for a finite time linear interaction quench is
studied in both large $\tau$ and small $\tau$ limits. In particular,  we reproduce  the results for the adiabatic and sudden quench limits derived earlier \cite{Dora_prl13}, and using perturbative solutions we estimate  the corrections  to both these
limiting situations in the early time limit as well as the infinite time limit. We also compare the perturbative results with the exact numerical ones, and obtain a reasonably good agreement between them. 

Let us summarize the interesting findings of our study: we find that the correction terms scales as $1/\tau$ in the large 
$\tau$ limit and as $\tau^2$ in the $\tau \to 0$  limit which show up both in the early time and the large time behaviour.
We propose generic scaling relations to justify the scaling of the correction terms. Finally, our results confirm that 
 for a finite $\tau$ the excited states contribute non-trivially to the echo even in the asymptotic limit ($t \to \infty$). 

Note that our results [see for {\it e.g.}, Eqs.~\eqref{LEexp1}-\eqref{aqtfLE}] depend on the non-universal short distance cutoff, $\alpha$. 
This may be a consequence of the fact that we are using a linear ramp, whose derivative has discontinuities at 
$t=0$ and $t=\tau$. A Fourier transform of such a non-analytic ramp has a fat high frequency tail, which is governed by a power law  \cite{DzairmagaPRB2011}. The high frequencies in the tail 
give rise to excitations that are beyond the low-energy description of the 
Luttinger model. We believe that if the linear ramp were replaced by a 
smooth quench protocol [{\it e.g.}, $\tanh (t/\tau)$-like], then the tail would be exponential and 
the inverse ultraviolet cutoff $\alpha$ would not appear in the results \footnote{This was pointed out by one of the referees.}.

Finally we note that the Luttinger model uses a linearized dispersion relation which is  valid only in a small window of energy around the Fermi points. In any realistic quenching scenario, the system will be excited to energies where, the non-linearities of the dispersion relation begin to play a role in the dynamics via the so called `umklapp' scattering of the right- and left-moving modes. 
 It is commonly argued in the literature that all these deviations are irrelevant in the renormalization group sense, which means that their effects are limited, or that their effects die out if we wait for long enough times. Another way to avoid exciting the system to very high energies (where the Luttinger model is not valid), is to tune the interaction parameter ($g$) in a regime which is much smaller than the Fermi energy of the system \cite{cazalila_prl06}. However, it is not completely clear, if the Luttinger model is appropriate to describe the quench dynamics in a realistic physical system, even though we believe that  the long-term dynamics of the system should be dominated by the low-energy excitations. 

\begin{acknowledgments}
 AA gratefully acknowledges funding from the INSPIRE faculty fellowship by DST  (Govt. of India), and from the Faculty Initiation Grant by IIT Kanpur. We thank the referees for their useful and constructive comments.
 \end{acknowledgments}

\end{document}